\documentclass[letter]{aa}  
\usepackage{graphicx}
\usepackage{txfonts}
\begin{document}

\title{Searching for companions down to 2 AU from $\beta$ Pictoris \\ using the $L'$-band AGPM coronagraph on VLT/NACO}

\author{O. Absil\inst{1}
        \and
        J. Milli\inst{2,3}
        \and
        D. Mawet\inst{2}
        \and
        A.-M. Lagrange\inst{3}
        \and
        J. Girard\inst{2}
        \and
        G. Chauvin\inst{3}
        \and \\
        A.~Boccaletti\inst{4}
        \and
        C.~Delacroix\inst{1}
        \and
        J.~Surdej\inst{1}
}

\institute{D\'epartement d'Astrophysique, G\'eophysique et Oc\'eanographie, Universit\'e de Li\`ege, All\'ee du Six Ao\^ut 17, B-4000 Li\`ege, Belgium
\and European Southern Observatory, Alonso de Cordova 3107, Casilla 19001, Vitacura, Santiago 19, Chile
\and UJF-Grenoble 1 / CNRS, Institut de Plan\'etologie et d'Astrophysique de Grenoble (IPAG) UMR 5274, F-38041 Grenoble, France
\and LESIA-Observatoire de Paris, CNRS, UPMC Univ.\ Paris 06, Univ.\ Paris-Diderot, 5 pl.~J.~Janssen, F-92195 Meudon, France
}

   \date{Received September 25, 2013; accepted November 6, 2013}

\abstract
{The orbit of the giant planet discovered around $\beta$~Pic is slightly inclined with respect to the outer parts of the debris disc, which creates a warp in the inner debris disc. This inclination might be explained by gravitational interactions with other planets.} 
{We aim to search for additional giant planets located at smaller angular separations from the star.}
{We used the new $L'$-band AGPM coronagraph on VLT/NACO, which provides an exquisite inner working angle. A long observing sequence was obtained on $\beta$~Pic in pupil-tracking mode. To derive sensitivity limits, the collected images were processed using a principal-component analysis technique specifically tailored to angular differential imaging.}
{No additional planet is detected down to an angular separation of $0\farcs2$ with a sensitivity better than $5M_{\rm Jup}$. Meaningful upper limits ($<10M_{\rm Jup}$) are derived down to an angular separation of $0\farcs1$, which corresponds to $2$~AU at the distance of $\beta$~Pic.}
{}

\keywords{Stars: planetary systems -- Stars: individual: $\beta$~Pic -- Techniques: high angular resolution}

\maketitle


\section{Introduction}

Historically, the first planetary system to be resolved around a main-sequence star was the debris disc imaged by \citet{Smith84} around $\beta$~Pic, a bright ($V=3.5$) and nearby ($d=19.3$~pc) A5V star. Since then, tremendous instrumental development has allowed the $\beta$~Pic planetary system to be studied with constantly improving angular resolution and dynamic range. This effort culminated with the discovery of a giant planet orbiting at about 8--9~AU from its host star \citep{Lagrange10}. This planet is now widely recognised to be responsible for the warp detected in the inner debris disc up to about 85~AU from the star, as first proposed by \citet{Mouillet97}. The origin of the dynamical excitation in the system, however, remains unclear. No other massive companion seems to be present at larger orbital distance, with upper limits of a few Jupiter masses \citep{Boccaletti09}. Additional planetary-mass objects were also searched for closer to the star using various observing techniques. So far, coronagraphic observations provided upper limits significantly higher than $10 M_{\rm Jup}$ in the $0\farcs2$ to $0\farcs3$ region \citep{Boccaletti09,Quanz10}, while interferometric observations ruled out the presence of objects with masses larger than about $50 M_{\rm Jup}$ within $0\farcs1$ \citep{Absil10}. Complementary information was also obtained with radial velocity measurements of the host star, providing detection limits for orbital periods up to 1000 days, that is, a maximum angular separation of about $0\farcs12$, with values in the planetary mass range for periods of up to 500 days \citep{Lagrange12}.

In this Letter, we aim to improve upon these previous studies and provide the best detection limits achieved so far at angular separations ranging from $0\farcs1$ to $0\farcs4$, using the new $L'$-band Annular Groove Phase Mask \citep[AGPM,][]{Mawet05} vector vortex coronagraph on VLT/NACO \citep{Mawet13}. The AGPM provides a compelling combination of an exquisite inner working angle (IWA), down to the diffraction limit of the telescope ($\sim 0\farcs1$ at $L'$ band), with an operating wavelength that is often regarded as a sweet spot for exoplanet imaging.


\section{Observations and data reduction}

\begin{figure*}[!t]
\begin{center}
\begin{tabular}{cc}
\includegraphics[width=7.5cm]{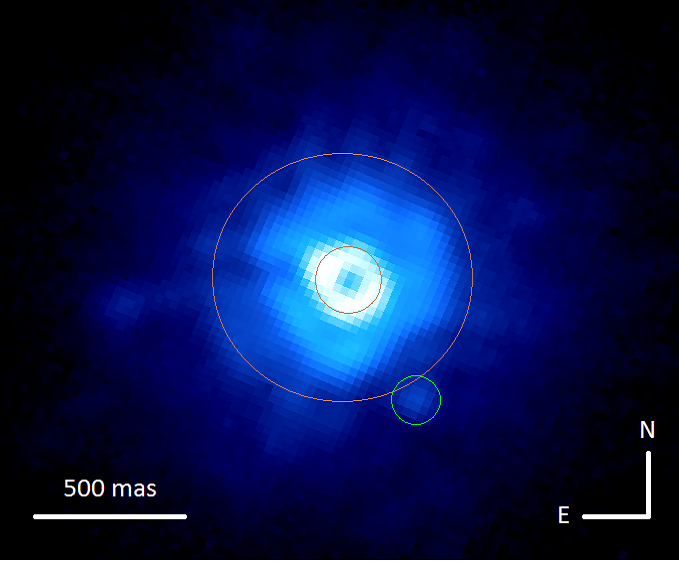} & \includegraphics[width=7.5cm]{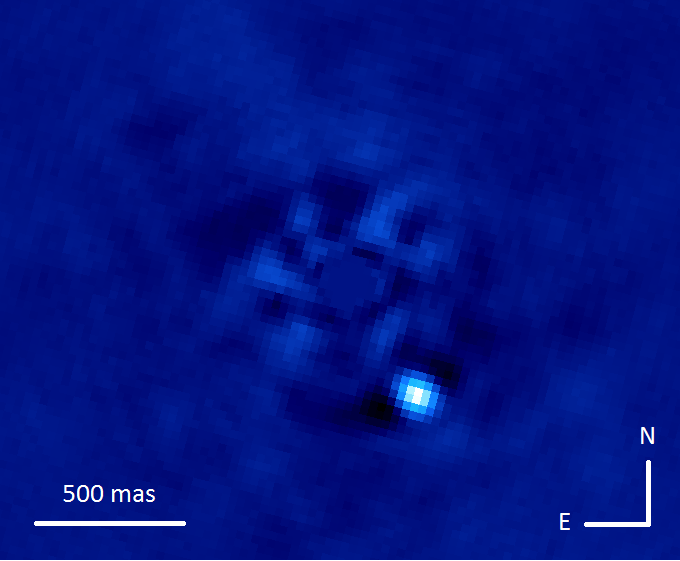}
\end{tabular}
\end{center}
\caption{\textit{Left.} Illustration of an individual image of the cleaned cube (log scale), showing that the companion can be readily identified without further image processing (cf.\ green circle). The region used for frame selection is located between the two brown circles. \textit{Right.} Final reduced image obtained as the median of the de-rotated sPCA-processed cleaned cube (linear scale). North is up and east to the left in both images.}
\label{fig:image}
\end{figure*}

On 31 January 2013, $\beta$ Pic was observed for about 3.5 hours at $L'$ band within the science verification observing run of NACO's new AGPM coronagraphic mode. The observations were obtained under fair seeing (${\rm FWHM} \sim 1\farcs0$), but the turbulence was fast ($\tau_0\simeq2$~msec), resulting in frequent openings of the AO loop despite the use of the infrared wave-front sensor. The Strehl ratio at $L'$ band typically ranged between 70\% and 75\% during the observations. Despite the mediocre observing conditions, $\beta$~Pic~b could be directly identified on the NACO real time display, thanks to the peak starlight extinction of about 50:1 provided by the AGPM (see Fig.~\ref{fig:image}, left). Individual observing blocks (OBs) consisted of 200 successive frames of 0.2 sec each, with the detector windowed to a $768\times768$ pixels region. Sequences of 10 or 20 OBs were obtained, each of them followed by three OBs of 50 frames on a nearby sky region and by re-centring the star under the AGPM mask to achieve maximal extinction. A total of 190 OBs were obtained on $\beta$~Pic, resulting in an on-source integration time of 114 min, out of an overall observing time of 204 min. Observations were taken in pupil-tracking mode, with a parallactic angle ranging from $-15\degr$ to $68\degr$. The undersized Lyot stop APO165 was used as in \citet{Mawet13}. Before and after the coronagraphic observations, we measured the non-coronagraphic point spread function (PSF) through the AGPM and the Lyot stop by placing the star away from the vortex centre. These observations also serve as photometric reference in our data analysis, where we take the airmass variation into account to properly normalise individual frames.

After applying basic cosmetic treatment to individual frames (flat-fielding, bad pixel/cosmic-ray removal), we started the data processing by performing frame selection to remove the frames affected by strong AO loop openings. We found that a convenient selection criterion is the standard deviation of the pixel intensity in the $1-4 \lambda/D$ region ($0\farcs1 - 0\farcs4$), which is most affected by residual starlight but does not contain the signal of $\beta$~Pic~b (located beyond $0\farcs4$ during our observations). To enhance the contribution of additional residual starlight in bad frames with respect to good ones, we performed a principal component analysis (PCA) of the whole image cube, and subtracted from each individual image its projection onto the first principal component (PC) before computing the standard deviation of the pixel intensity. A histogram of all measured standard deviations was built, and the threshold for frame selection was set at a $1\sigma$ level above the median. The fraction of frames rejected by this selection process was about $15\%$. We found this image-processing step to be critical in reaching the best possible image quality.

The second image-processing step consisted of accurately recentring the frames. We performed a (negative) Gaussian fit to the PSF centre, which resembles a dark hole surrounded by a bright doughnut (see Fig.~\ref{fig:image}, left). This shape is mainly due to the combination of residual tip-tilt with the off-axis transmission profile of the AGPM. By recentring each image using the fitted Gaussian, we made sure that the centre of AGPM is placed at the exact same position in all individual frames, with a typical accuracy of 0.005 pixel (i.e., about 0.1~mas). We emphasize that this centring accuracy pertains to the position of the AGPM, not of the star itself. We then subtracted from each frame the estimated contribution of the sky based on the median of neighbouring sky observations. A new image cube was then created by averaging 40 successive frames (i.e., 8~sec of effective integration time), resulting in 612 individual images in the cleaned, recentred cube.

Finally, we used our implementation of the KLIP algorithm \citep{Soummer12} to produce a final image of the $\beta$~Pic system based on the cleaned cube, taking advantage of angular differential imaging (ADI). In the KLIP algorithm, the whole ADI image sequence is used as a PSF library, to which a PCA treatment is applied. In the presence of an off-axis companion, the PC computed on the ADI cube are expected to contain (part of) the companion signal. To prevent the planet from being partly removed from the individual images when subtracting its projection onto the first $K_{\rm klip}$ PC, we decided to implement a ``smart'' version of the PCA (or \emph{sPCA}), where the image library is built only from images where the off-axis companion has rotated by $1\lambda/D$ or more with respect to the image under consideration. In this way, the PC will not contain any (or only a very small amount of) signal from the companion at its current position. At the angular separation of $\beta$~Pic~b ($\sim 0\farcs45$), this translates into rejecting from the PSF library all images that have parallactic angles within about $15\degr$ from the image under consideration. We performed the sPCA in a region of about $3\arcsec$ in radius around the star, after masking out the region located within the IWA of the AGPM \citep[$\sim \lambda/D$,][]{Delacroix13}. The final image, obtained as the median of the de-rotated cube after sPCA processing, is shown in Fig.~\ref{fig:image}. The number of PC kept in this analysis is $K_{\rm klip}=30$ (out of 612), which is a good compromise to prominently reveal the planetary companion. The black spots on either sides of the planet are artefacts related to the rotation of the planet around the optical axis in the image sequence taken in pupil-tracking mode.


\section{Analysis of $\beta$ Pic b}

	\subsection{Photometry}

Even when using sPCA, part of the planetary signal is self-subtracted during the stellar halo removal process. To retrieve the photometric information without bias, we used the negative fake companion technique \citep{Lagrange10,Bonnefoy11}. The method proceeds as follows: (i) estimate the (biased) position and flux of the companion from the first reduced image; (ii) use the measured off-axis PSF as a template to remove this first estimate from the cleaned data cube before applying PCA; and (iii) iterate on the position ($x$, $y$) and flux until a well-chosen figure of merit -- here, the weighted sum of the squared pixel intensity in a pie chart aperture centred on the first estimate of the companion position, $2.44\lambda/D$ in radius and $6 \times 1.22\lambda/D$ in azimuth -- is minimized. The minimization was performed with the simplex-amoeba optimization. An exploration of the figure of merit (equivalent to a $\chi^2$) around the best-fit position is used to evaluate the statistical error bar on the photometry of the planet ($0.15$~mag at $1\sigma$ assuming Gaussian noise). Adding the contribution of photometric variations ($0.05$~mag for a photometric night) to the error bar, and correcting the companion photometry for the off-axis transmission profile of the AGPM \citep{Delacroix13}, we obtain a final contrast $\Delta L' = 8.01 \pm 0.16$~mag between the planet and the star. This result agrees within error bars with the contrast ($7.8 \pm 0.3$~mag) found by \citet{Lagrange10} in the same photometric band. The final photometry of $\beta$~Pic~b is given in Table~\ref{tab:planet}. With an absolute magnitude $M_{L'}=10.02\pm0.16$, and assuming an age of $12^{+8}_{-4}$~Myr for the system as in \citet{Bonnefoy13}, the estimated mass of $\beta$~Pic~b amounts to $8.0^{+3.2}_{-2.1} M_{\rm Jup}$, based on the BT-Settl models of \citet{Allard11}. The recent revision of the age \citep[$21\pm4$~Myr,][]{Binks13} would lead to a mass of $10.6^{+1.2}_{-1.8} M_{\rm Jup}$.

\begin{table}
\caption{Measured properties of the $\beta$~Pic system}
\label{tab:planet}
\centering
\begin{tabular}{c c c}
\hline\hline
Parameter & $\beta$ Pic A & $\beta$ Pic b \\
\hline
$d$ (pc) & $19.44 \pm 0.05$ & --- \\
$L'$ (mag) & $3.454\pm0.003$ & $11.46\pm0.16$ \\
$M_{L'}$ (mag) & $2.011\pm0.006$ & $10.02 \pm 0.16$ \\
Separation (mas) & --- & $452 \pm 10$ \\
PA (deg) & --- & $211.2 \pm 1.3$ \\
\hline
\end{tabular}
\end{table}

	\subsection{Astrometry}

The astrometry of the companion was evaluated using the negative fake companion technique, following the same procedure as for the photometry. The main limitation in the astrometric accuracy in high-contrast imaging generally comes from the imperfect knowledge of the star position in the images (either because it is masked by a coronagraph, or because the star is saturated in the images). In our case, the recentring procedure allowed us to make sure that the AGPM position is the same in all images with an accuracy of about 0.1~mas. This does not mean that the position of the star in the image is known with such an accuracy, however, as it is generally not perfectly centred on the AGPM. We can evaluate the misalignment between the star and the AGPM centre thanks to the instantaneous starlight rejection rate. Assuming that the residual starlight is entirely due to misalignment, we derive a mean upper limit of about 8.5~mas on the error on the position of the star with respect to the AGPM centre. This error is folded into our astrometric error budget. Although this error bar is not drastically improved with respect to saturated $L'$-band data sets \citep{Chauvin12}, the more robust knowledge of the stellar position is still a clear advantage of the AGPM observing mode.

The other contributors to the astrometric errors are the statistical error on the companion position and the calibration errors. The statistical error bar is estimated by exploring the figure of merit around the best-fit solution obtained with the negative fake companion technique, yielding a $1\sigma$ error of 4.5~mas. The calibration errors, related to the platescale, true north orientation and uncertainty on the NACO rotator offset, are computed as in \citet{Chauvin12}, using an estimate platescale of $27.10 \pm 0.04$~mas, a true north of $-0\fdg45 \pm 0\fdg09$, and a rotator offset of $104\fdg84 \pm 0\fdg01$. They are negligible in the final error budget. The final astrometry with error bars is given in Table~\ref{tab:planet}. A comparison with the orbital position predictions of \citet{Chauvin12} suggests that the planet is now starting to recess.


\section{Search for additional, inner companions}

In this section, we exploit the exquisite IWA of the AGPM to provide new constraints on the presence of planetary companions down to an angular distance of $1\lambda/D$ ($0\farcs1$) at L' band. At the distance of $\beta$~Pic, this represents a linear separation of about 2~AU. To derive unbiased contrast curves, we removed the contribution of $\beta$~Pic~b from all individual images in the cube, using the best-fit photometry and astrometry found in the previous section and the off-axis PSF obtained during our observations. The noise level can then be computed as the standard deviation of the pixel intensity in concentric annuli, or equivalently as the azimuthal median of the noise computed locally in small square boxes, even at the location of the (removed) planet. However, the final contrast curve cannot be trusted within a circular region about $1.22 \lambda/D$ ($0\farcs13$) in radius around the position of $\beta$~Pic~b, because the presence of an additional companion within this zone would result in a partial PSF overlap and in an improper subtraction of $\beta$~Pic~b from the image cube.

We implemented an annulus-wise version of the sPCA method, where the exclusion criterion on the parallactic angles to be included in the PSF library is computed separately in thin annuli $2\lambda/D$ in width. To assess the amount of self-subtraction of potential companions at any given angular separation from the star, we introduced fake companions in our cube, separated by a few $\lambda/D$ from each other and placed on three radial branches separated by $120\degr$ in azimuth. The fake companions were injected at $20\sigma$ above the noise computed after a first pass of the sPCA algorithm on the cube without fake companions. By measuring the photometry of the fake companions in the final reduced image after a second pass of the sPCA algorithm and comparing it to their input flux, we inferred the attenuation of the sPCA algorithm in ADI mode. As expected, the self-subtraction strongly depends on the width of the exclusion zone in terms of parallactic angle, but the final contrast curve is only weakly affected by this parameter. Taking self-subtraction into account, small exclusion zones provide slightly better detection limits at very small angular separations, because the frames that are more correlated (i.e., closer in time) to the current frame are then kept in the library. An exclusion zone of only $0.1\lambda/D$ is used in the following. With this criterion, and taking $K_{\rm klip}=20$ in each ring, the companion self-subtraction is significant in the innermost parts of the search region, with only about 40\% (resp.\ 20\%) of the signal making it through at $0\farcs5$ (resp.\ $0\farcs25$).

\begin{figure}[!t]
\begin{center}
\includegraphics[width=9cm]{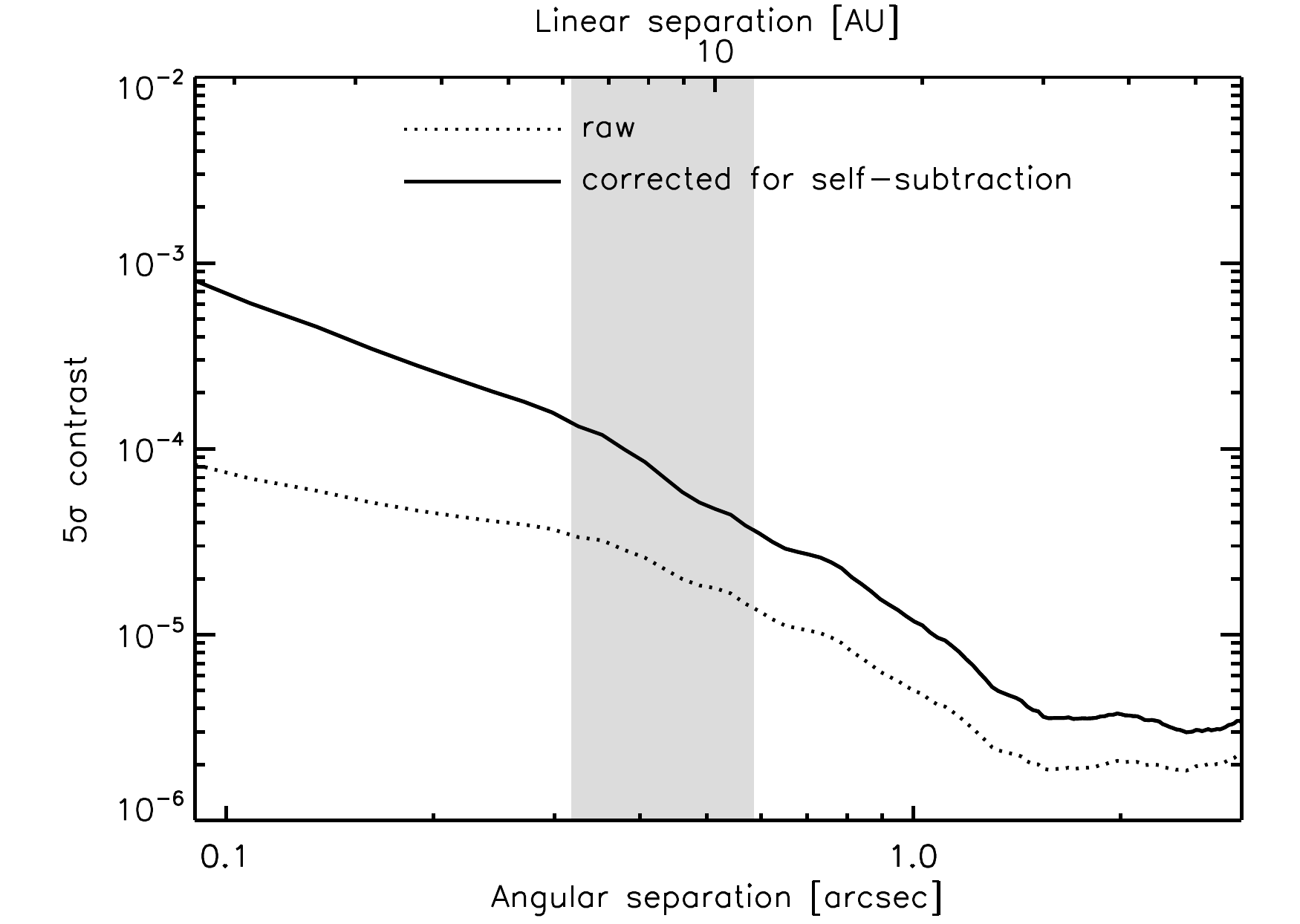}
\end{center}
\caption{$5\sigma$ detectability limits in terms of contrast for point-like companions around $\beta$~Pic (solid line). The dotted line shows the contrast curve that would be derived if self-subtraction encountered in the ADI-PCA data processing was not taken into account. The grey-shaded region recalls that the detection limits are not valid in a small region of $\sim 0\farcs13$ in radius around $\beta$~Pic~b.}
\label{fig:contrast}
\end{figure}

The final contrast curves, computed as five times the standard deviation of the pixel intensities in $\lambda/D$-wide annuli after applying a median filter on a $\lambda/D$-wide moving box, are displayed in Fig.~\ref{fig:contrast} before and after taking into account companion self-subtraction. Owing to the small number of independent resolution elements in $\lambda/D$-wide annuli at short angular separations, the confidence level ($1-3\times 10^{-7}$) associated to a $5\sigma$ detection limit for pure Gaussian noise is not preserved, as discussed in Mawet et al.\ (in prep). Reducing the confidence level (i.e., increasing the false alarm probability) at small angles is, however, not a severe limitation to the validity of the detection limits, not only because no candidate companion is found in our case, but also because one can readily distinguish false positives (bright speckles) from true companions at a few $\lambda/D$ with follow-up observations (background point-like sources being very unlikely in such a small search region). 

The detection limits presented in Fig.~\ref{fig:contrast} are based on the debatable assumption of Gaussian noise. To address this point, which was shown by \citet{Marois08} to be a potential source of bias on the sensitivity limits, we constructed the histogram of the pixel intensities in the final image (the median of the de-rotated individual sPCA-processed images in the cleaned cube), using annuli $\lambda/D$ in width. Using the Shapiro-Wilk test \citep{Shapiro65}, we verified that for most annuli within the search region, the statistics of the pixel intensity can be considered as Gaussian, with p-values typically ranging from 10\% to 90\%.

\begin{figure}[!t]
\begin{center}
\includegraphics[width=9cm]{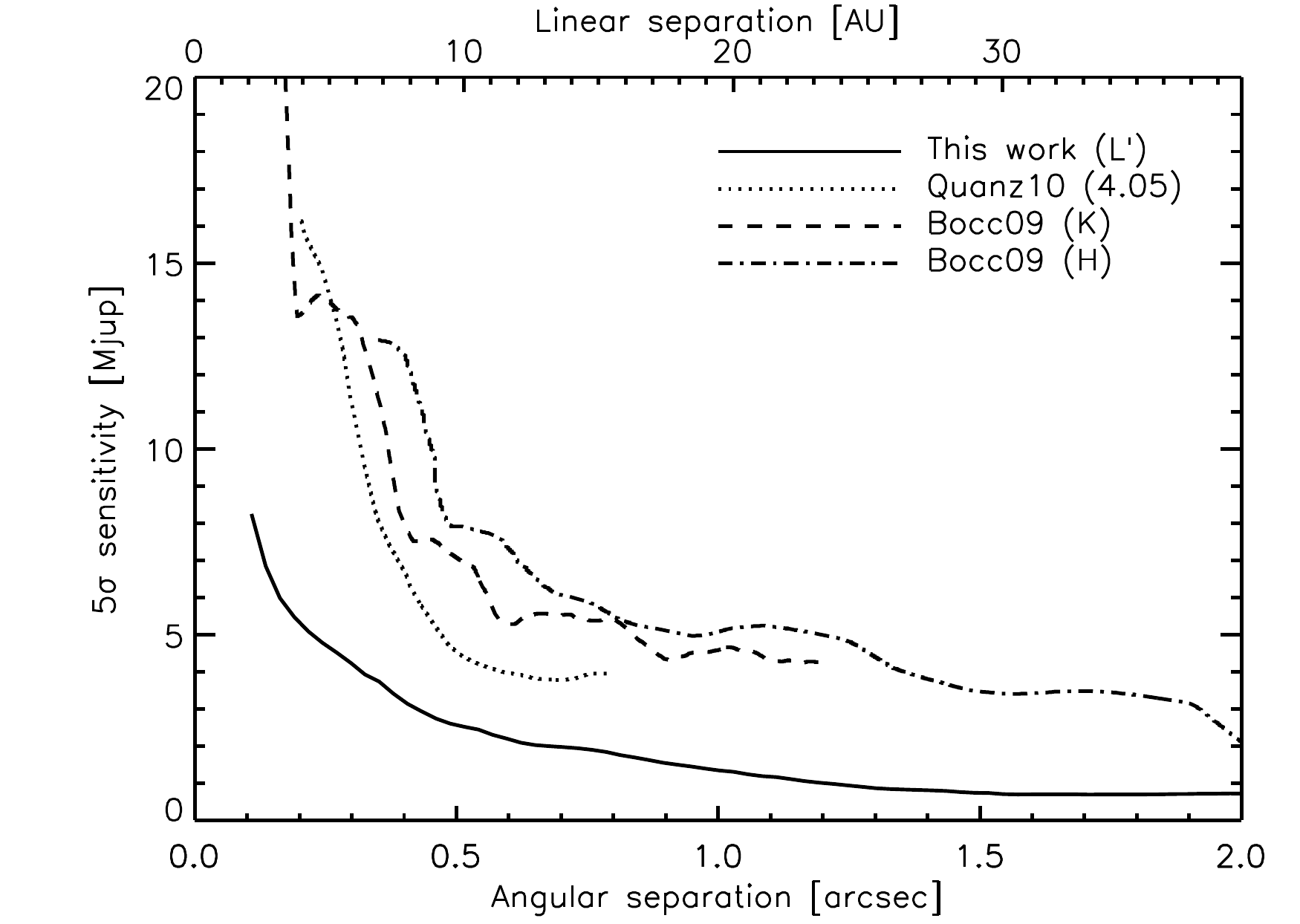}
\end{center}
\caption{Sensitivity limits in terms of mass derived from our data set (solid line), compared with the results of \citet[dashed and dashed-dotted]{Boccaletti09} and \citet[dotted]{Quanz10} using the same evolutionary models (see text). This figure cannot serve as a direct comparison between the sensitivity of these NACO observing modes, as ADI was not used in the three previously published data sets.}
\label{fig:detlim}
\end{figure}

To convert our detection limits into sensitivity limits in terms of planetary mass, we used the BT-Settl models of \citet{Allard11}. The result is given as the solid curve in Fig.~\ref{fig:detlim}, where the sensitivity limits of previous high-contrast imaging studies have been plotted for comparison using the same BT-Settl models. This figure illustrates the high gain in sensitivity at short angular separation enabled by the $L'$-band AGPM, which allows planets more massive than $5M_{\rm Jup}$ to be ruled out down to about $0\farcs2$ from $\beta$~Pic. Detection limits within the planetary-mass regime are derived down to $0\farcs1$, although the strong self-subtraction, the bright speckles, and the small number of independent resolution elements urge us to take the detection limit with caution in the $0\farcs1 - 0\farcs2$ region. The sensitivity to companions located farther away than $\beta$~Pic~b is also excellent, down to about $1M_{\rm Jup}$ beyond $1\farcs5$. With 768 pixels in our images, the outer working angle is about $10\arcsec$.

\section{Concluding remarks}

We have presented the results of our search for additional planetary companions around $\beta$~Pic. These results nicely complement those obtained at shorter orbital periods  using radial velocity measurements \citep{Lagrange12}. Combined, these two studies can now exclude the presence of giant planets with masses similar to that of $\beta$~Pic~b at most orbital distances (only in the 1--2~AU region are the detection limits still in the 10--20\,$M_{\rm Jup}$ range). The fact that we do not detect any additional, massive companion around $\beta$~Pic agrees with the conclusion that $\beta$~Pic~b is the sole body responsible for the warp in the inner debris disc, as proposed by \citet{Lagrange12}. It suggests that the dynamical excitation in the $\beta$~Pic inner system is most probably not directly related to gravitational interactions with other massive bodies ($>5M_{\rm Jup}$) located in the inner system.

The excellent sensitivity limits that we obtained down to very short angular separations from $\beta$~Pic illustrates the potential of the $L'$-band AGPM coronagraph recently installed on NACO, despite the mediocre seeing conditions. We expect that the NACO/AGPM sensitivity can be further improved under better adaptive optics (AO) correction, with an improved calibration of static instrumental aberrations, and with an optimal Lyot stop. The AGPM mode on NACO would be very well suited to follow-up and characterise new companions found by upcoming near-infrared imagers aided by extreme AO systems. Although our observing sequence on $\beta$~Pic spans a wide range of parallactic angles ($83\degr$), we note that the sensitivity limit close to the IWA is still largely affected by the limited displacement of any candidate companion between individual images obtained in pupil tracking mode. We suggest that reference-star differential imaging, an observing strategy mostly dropped in favour of ADI during the past few years, may be the most appropriate way to unleash the full potential of phase masks such as the AGPM in terms of IWA, when operating under good and stable AO correction.


\begin{acknowledgements}
The research leading to these results has received funding from the European Research Council
under the FP7 through ERC Starting Grant Agreement no.\ 337569, and from the Communaut\'e fran\c caise de Belgique -- Actions de recherche concert\'ees -- Acad\'emie universitaire Wallonie-Europe.
\end{acknowledgements}

\bibliographystyle{aa} 
\bibliography{betPic_AGPM} 

\end{document}